\documentstyle[12pt]{article}
\begin{document}

\thispagestyle{empty}

\baselineskip=0.6cm

\noindent P.~N.~Lebedev Physical Institute              \hfill
Preprint FIAN/TD/4--94\\ I.~E.~Tamm Theory Department       \hfill
\begin{flushright}{April 1994}\end{flushright}

\begin{center}

\vspace{0.5in}

{\Large\bf ON THE MULTILEVEL FIELD--ANTIFIELD }

\bigskip

\vspace{0.41cm}

{\Large\bf FORMALISM WITH THE MOST GENERAL}

\bigskip

\vspace{0.41cm}

{\Large\bf LAGRANGIAN HYPERGAUGES}

\bigskip

\vspace{0.3in}

{\large  I.~A.~Batalin and I.~V.~Tyutin}\\
\medskip  {\it Department of Theoretical Physics} \\ {\it  P.~N.~Lebedev
Physical Institute} \\ {\it Leninsky prospect, 53, 117 924, Moscow,
Russia}$^{\dagger}$\\

\end{center}

\vspace{1.5cm}

\centerline{\bf ABSTRACT}
\begin{quotation}

The multilevel field--antifield formalism is constructed in a geometrically
covariant way without imposing the unimodularity conditions on the hypergauge
functions. Thus the previously given version [1,2] is extended to cover the
most general case of Lagrangian surface bases. It is shown that the extra
measure factors, required to enter the gauge--independent functional
integrals, can be included naturally into the multilevel scheme by modifying
the boundary conditions to the quantum master equation.

\end{quotation}

\vfill

\noindent

$^{\dagger}$ E-mail address: batalin@fian.free.net,tyutin@fian.free.net

\newpage

\setcounter{page}{2}

\section{Introduction}

In Refs. [1,2] we have suggested a geometrically covariant multilevel
generalization of the field--antifield BV formalism [3 -- 5]. As is well known,
the field-antifield formalism is nothing else but the universal hypergauge
theory whose action is determined by the quantum  master equation, and the
corresponding hypergauge generators are nilpotent.

To eliminate the characteristic degeneracy  of the universal BV theory,  one
needs the hypergauge conditions singling out a Lagrangian
surface in the field--antifield phase space. By definition, the surface
$G_a=0$ is called the Lagrangian one if the antibracket involution relations
$(G_a,G_b)=G_cU^c_{ab}$ are satisfied. However, in Refs. [1,2] we have imposed
stronger restriction on the hypergauge functions $G_a$ by requiring the
unimodularity conditions $\Delta G_a-U^b_{ba}(-1)^{\varepsilon_b}=G_bV^b_a$ to
be satisfied too, where $\Delta$ is the well known nilpotent operator entering
the quantum master equation $\Delta\exp({\imath\over\hbar}W)=0$. As it has
been explained in Refs. [1,2], the unimodularity conditions do not restrict
the arbitrariness of an intrinsic gauge fermion encoded by means of the
functions $G_a$. Nevertheless, these extra conditions restrict rather strongly
the possibilities to change the Lagrangian surface basis by making linear
transformations of the form $G_a\rightarrow G_b\Lambda^b_a$. Being the
unimodularity conditions fulfilled, the Jacobian of the $\delta$--functions
$\delta(G)$ is constant, and the admitted generators $\delta\Lambda^a_b$ of the
infinitesimal basis transformations,
$\Lambda^a_b=\delta^a_b+\delta\Lambda^a_b$,
should be supertraceless : $(-1)^{\varepsilon_a}\delta\Lambda^a_a=0$.

In the present paper we cancel the unimodularity conditions and thereby extend
the construction of Refs. [1,2] to cover the most general case of the
Lagrangian surface bases. Thus the antibracket involution relations and usual
admissibility conditions are the only restrictions imposed on the hypergauge
functions.

In order to provide for a gauge independence to the  functional integral, we
insert the extra measure factor into the integrand, and then convert  former
unimodularity conditions into the equations determining this extra measure to
absorb the  transformation  of  the hypergauge $\delta$--function Jacobian.

As a result, it will be shown that all the extra measure factors can be
included
naturally into the multilevel scheme by modifying the boundary conditions to
the
quantum master equation.

As usual, we denote by $\varepsilon(A)$ the Grassmann parity of a quantity
$A$, and $\hbox{Ber}K$ stands for the Berezinian (superdeterminant) of a
supermatrix $K$.

\section{A Modified Version to the First Level Formalism}

Let $\Gamma^A$, $A=1,\ldots,2N$, $\varepsilon(\Gamma^A)\equiv\varepsilon_A$,
be a total set of field--antifield variables coordinatizing the original phase
space locally.

We define the antisymplectic differential $\Delta$ to be a general
second--order fermionic operator without the derivativeless term,

$$\Delta={1\over2}(-1)^{\varepsilon_A}M^{-1}\partial_AME^{AB}\partial_B,
\eqno{(2.1)}$$
required to satisfy the nilpotency condition, $\Delta^2=0$, so that
$E^{AB}(\Gamma)$ appears to be antisymplectic metric satisfying the Jacobi
identity and thus yielding the antibracket operation

$$(F,G)\equiv F\overleftarrow{\partial_A}E^{AB}\overrightarrow{\partial_B}G
{}.
\eqno{(2.2)}$$

Let us remind here the  corresponding  differentiation formulae for the
antibracket (2.2) and ordinary product $FG$ :

$$\Delta(F,G)=(\Delta F,G)+(F,\Delta G)(-1)^{\varepsilon(F)+1},
\eqno{(2.3)}$$

$$\Delta(FG)=(\Delta F)G+(F,G)(-1)^{\varepsilon(F)}+
F(\Delta G)(-1)^{\varepsilon(F)}. \eqno{(2.4)}$$

A modified version to the first level functional integral is defined as
follows :

$$Z=\int\!\!\exp\{{i\over\hbar}[W(\Gamma;\hbar)+G_a(\Gamma;\hbar)\pi^a]
-H(\Gamma;\hbar)\}d\mu, \eqno{(2.5)}$$
where

$$
d\mu=Md\Gamma d\pi
\eqno{(2.6)}$$
is the integration measure, the action $W(\Gamma;\hbar)$ satisfies the quantum
master equation

$$\Delta\exp\{{i\over\hbar}W(\Gamma;\hbar)\}=0, \eqno{(2.7)}$$
$\pi^a, a=1,\ldots,N$, $\varepsilon(\pi^a)\equiv\varepsilon_a$, are  the
Lagrangian  multipliers introducing the hypergauge functions $G_a$ that
satisfy the general involution relations

$$(G_a,G_b)=G_cU^c_{ab} \eqno{(2.8)}$$
with some structure coefficients $U^c_{ab}(\Gamma;\hbar)$, the  function
$H(\Gamma;\hbar)$ satisfies the equations

$$(H,G_a)=\Delta G_a-U^b_{ba}(-1)^{\varepsilon_b}-G_bV^b_a \eqno{(2.9)}$$
with some structure coefficients $V^b_a(\Gamma;\hbar)$. It follows from (2.9)
that

$$\Delta H-{1\over2}(H,H)+V^a_a=G_a\tilde{G}{}^a \eqno{(2.10)}$$
with $\tilde{G}{}^a(\Gamma;\hbar)$ to be some functions of the original phase
variables.

As compared with the previously given version [1,2] of the first level
formalism, the modified functional integral (2.5) is the most generalization in
what concerns the choice  of hypergauge functions $G_a$. The only restriction,
imposed on $G_a$ in the present version, is that these functions should satisfy
the  general involution relations (2.8). That means that the hypergauge
functions $G_a$ describe, from the purely geometric viewpoint, the most general
basis of the Lagrangian surface $G_a=0$.

In order to provide for a gauge independence to the modified functional
integral (2.5), one should insert the extra measure factor $\exp(-H)$ into the
integrand of (2.5), and then convert former unimodularity conditions into the
equations (2.9) determining the function $H$ under the only conditions
(2.8) imposed on $G_a$. It is evident that the modified version (2.5)
coincides with the previously given one if the r.h.s. of (2.9) vanishes.

Now, let us consider the gauge independence in more details. First of, all we
observe that the integrand of (2.5) is invariant under the generalized
BRST-type
transformations:

$$\delta\Gamma^A=(\Gamma^A,-W+G_a\pi^a+\imath\hbar H)\mu, \eqno{(2.11)}$$

$$\delta\pi^a=(-U^a_{bc}\pi^c\pi^b(-1)^{\varepsilon_b}+2i\hbar V^a_b\pi^b+
2(i\hbar)^2\tilde{G}{}^a)\mu,\eqno{(2.12)}$$
where $\mu=\hbox{const}$, $\varepsilon(\mu)=1$.

Choosing the parameter $\mu$ to be an arbitrary function

$$\mu={i\over2\hbar}\delta X(\Gamma), \eqno{(2.13)}$$
and making the additional variations

$$\delta\Gamma^A={1\over2}(\Gamma^A,\delta X),\quad\delta\pi^a=
\delta\Lambda^a_b\pi^b, \eqno{(2.14)}$$
with arbitrary functions $\delta\Lambda^a_b(\Gamma)$, one generates the
following effective changes in the integrand of (2.5) :

$$
\delta G_a=(G_a,\delta X)+G_b\delta\Lambda^b_a,
\eqno{(2.15)}$$

$$
\delta H=-\Delta\delta X+(H,\delta X)-(-1)^{\varepsilon_a}\delta\Lambda^a_a,
\eqno{(2.16)}$$
where the first and second terms in the r.h.s. of (2.15) describe,
respectively,
the most general changes of the hipergauge surface and its basis.

On the other hand, the transformations (2.15) retain the form of the equations
(2.8), (2.9) by inducing the following variations of structure coefficients :

$$
\delta U^a_{bc}=(U^a_{bc},\delta X)+[(\delta\Lambda^a_b,G_c)+
U^a_{bd}\delta\Lambda^d_c]-[(\delta\Lambda^a_c,G_b)+U^a_{cd}\delta\Lambda^d_b]
(-1)^{(\varepsilon_b+1)(\varepsilon_c+1)}-\delta\Lambda^a_dU^d_{bc},
\eqno{(2.17)}$$

$$
\delta V^a_b=(V^a_b,\delta X)+V^a_d\delta\Lambda^d_b-\delta\Lambda^a_dV^d_b+
[\Delta\delta\Lambda^a_b-(H,\delta\Lambda^a_b)](-1)^{\varepsilon_a}.
\eqno{(2.18)}$$

Due to the arguments analogous to the ones given in Refs. [1,2] we conclude
that
the modified functional integral (2.5) does not depend on hypergauge fixing.

\section{A Modified Version to the $n$-th Level Formalism}

In this Section we construct inductively the $n$-th level modified functional
integral for $n=2,3,\ldots$.

First, let us define recursively the $n$-th level set of variables of the
field--antifield phase space

$$\Gamma^{(n)A_{(n)}}\equiv\{\Gamma^{(n-1)A_{(n-1)}}; \pi^{(n-1)a},
\pi^{*(n-1)}_a\}, \eqno{(3.1)}$$
where

$$\Gamma^{(1)A_{(1)}}\equiv\Gamma^A,\quad\pi^{(1)a}\equiv\pi^a. \eqno{(3.2)}$$
with $\pi^{(n-1)a}$ and $\pi^{*(n-1)}_a$ to be the $(n-1)$-th level
Lagrangian multipliers and their conjugated antifields, respectively, so that

$$\varepsilon(\pi^{(n)a})=\varepsilon(\pi^{*(n)}_a)+1=\varepsilon_a
+n-1. \eqno{(3.3)}$$

In what follows all the antibrackets, ( , ), are understood to include the
totally extended set (3.2) of field--antifield variables, and the only nonzero
elementary antibrackets for the Lagrangian multipliers are

$$
(\pi^{(n)a},\pi^{(n)}_b)=\delta^{(m)(n)}\delta^a_b.
\eqno{(3.4)}$$

Further, one constructs recursively the nilpotent operators $\Delta^{(n)}$ :

$$\Delta^{(n)}\equiv\Delta^{(n-1)}+(-1)^{(\varepsilon_a+n)}
{\partial\over\partial\pi^{(n-1)a}}{\partial\over\partial\pi^{*(n-1)}_a},
\eqno{(3.5)}$$

$$\Delta^{(1)}\equiv\Delta. \eqno{(3.6)}$$

Let us assign to the $n$-th level, $n\ge2$, the corresponding Planck constant
$\hbar^{(n)}$, $\varepsilon(\hbar^{(n)})=0$, in addition to the usual one
$\hbar$, together with the new quantum number called the Planck parity
$\hbox{Pl}^{(n)}$:

$$\hbox{Pl}^{(n)}(\Gamma^{(n-1)})=\hbox{Pl}^{(n)}(\hbar)=0, \eqno{(3.7)}$$

$$\hbox{Pl}^{(n)}(\hbar^{(n)})=\hbox{Pl}^{(n)}(\pi^{(n-1)})=
-\hbox{Pl}^{(n)}(\pi^{*(n-1)})=1. \eqno{(3.8)}$$

The $n$-th level quantum action $W^{(n)}(\Gamma^{(n)};\hbar;\hbar^{(n)})$ is
defined to satisfy the quantum master equation:

$$\Delta^{(n)}\exp\{{\imath\over\hbar^{(n)}}
W^{(n)}(\Gamma^{(n)};\hbar;\hbar^{(n)})\}=0. \eqno{(3.9)}$$

The action $W^{(n)}$ possesses the quantum numbers:

$$\varepsilon(W^{(n)}(\Gamma^{(n)};\hbar;\hbar^{(n)}))=0,\quad
\hbox{Pl}^{(n)}(W^{(n)}(\Gamma^{(n)};\hbar;\hbar^{(n)}))=1, \eqno{(3.10)}$$
and has the following series expansion in powers of $\hbar^{(n)}$,
$\pi^{(n-1)}$, $\pi^{*(n-1)}$ :

$$W^{(n)}(\Gamma^{(n)};\hbar;\hbar^{(n)})=
\Omega^{(n)}(\Gamma^{(n)};\hbar)+\imath\hbar^{(n)}
\Xi^{(n)}(\Gamma^{(n)};\hbar)+
(\imath\hbar^{(n)})^2\tilde{\Omega}{}^{(n)}(\Gamma^{(n)};\hbar)
+\ldots, \eqno{(3.11)}$$

$$\begin{array}{c}
\Omega^{(n)}(\Gamma^{(n)};\hbar)=G_a^{(n-1)}(\Gamma^{(n-1)};\hbar)
\pi^{(n-1)a}+ \\[9pt]
+{1\over2}\pi^{*(n-1)}_cU_{ab}^{(n-1)c}(\Gamma^{(n-1)};\hbar)
\pi^{(n-1)b}\pi^{(n-1)a}(-1)^{(\varepsilon_a+n)}+\ldots,
\end{array} \eqno{(3.12)}$$

$$\Xi^{(n)}(\Gamma^{(n)};\hbar)=H^{(n-1)}(\Gamma^{(n-1)};\hbar)
+\pi^{*(n-1)}_aV_b^{(n-1)a}(\Gamma^{(n-1)};\hbar)\pi^{(n-1)b}+\ldots,
\eqno{(3.13)}$$

$$\tilde{\Omega}^{(n)}(\Gamma^{(n)};\hbar)=\pi^{*(n-1)}_a
\tilde{G}^{(n-1)a}(\Gamma^{(n-1)};\hbar)+\ldots, \eqno{(3.14)}$$

Substituting the expansion (3.11) for $W^{(n)}$ into the quantum master
equation (3.9), we find the following equations for the functions
$\Omega^{(n)}$, $\Xi^{(n)}$, $\tilde{\Omega}{}^{(n)}$, $n\ge2$ :

$$(\Omega^{(n)},\Omega^{(n)})=0, \eqno{(3.15)}$$

$$(\Omega^{(n)},\Xi^{(n)})=\Delta\hspace{-0,05cm}^{(n)}\hspace{0.05cm}
\Omega^{(n)}, \eqno{(3.16)}$$

$$(\Omega^{(n)},\tilde{\Omega}{}^{(n)})=
\Delta\hspace{-0,05cm}^{(n)}\hspace{0.05cm}\Xi^{(n)}-
{1\over2}(\Xi^{(n)},\Xi^{(n)}). \eqno{(3.17)}$$

To the lowest orders in $\pi^{(n-1)}$, $\pi^{*(n-1)}$ these equations give:

$$(G^{(n-1)}_a,G^{(n-1)}_b)=G^{(n-1)}_cU^{(n-1)c}_{ab}, \eqno{(3.18)}$$

$$(H^{(n-1)},G^{(n-1)}_a)=\Delta^{(n-1)}G^{(n-1)}_a+
U^{(n-1)b}_{ba}(-1)^{(\varepsilon_b+n-1)}-
G^{(n-1)}_bV^{(n-1)b}_a, \eqno{(3.19)}$$

$$\Delta^{(n-1)}H^{(n-1)}-{1\over2}(H^{(n-1)},H^{(n-1)})+
V^{(n-1)a}_a=G^{(n-1)}_a\tilde{G}{}^{(n-1)a}, \eqno{(3.20)}$$

The general involution relations (3.18) and usual gauge admissibility
conditions are the only restrictions imposed on the hypergauge functions
$G^{(n-1)}_a(\Gamma^{(n-1)};\hbar)$.

At $n=2$ we identify in (3.18) -- (3.20) :

$$G^{(1)}_a\equiv G_a,\quad U^{(1)c}_{ab}\equiv U^c_{ab}, \eqno{(3.21)}$$

$$H^{(1)}\equiv H,\quad V^{(1)b}_a\equiv V^b_a,\quad
\tilde{G}{}^{(1)a}\equiv\tilde{G}{}^a. \eqno{(3.22)}$$
At $n>2$ the equation (3.20) is certainly compatible with the one (3.19) but
already does not follow from the latter.  We  consider the equations (3.19),
(3.20) at $n>2$ to determine $H^{(n-1)}$ under the boundary conditions

$$H^{(n-1)}(\Gamma;\hbar)\big|_{G^{(n-1)}=0,\pi^{(n-2)}=0}=
-{\imath\over\hbar}W^{(n-2)}(\Gamma^{(n-2)};\hbar;\hbar)-
\ln\hbox{Ber}\bigl(\pi^{(n-2)},G^{(n-1)}(\Gamma^{(n-2)};\hbar)\bigr),
\eqno{(3.23)}$$
where we identify at $n=3$ :

$$W^{(1)}(\Gamma^{(1)};\hbar;\hbar)\equiv W(\Gamma;\hbar). \eqno{(3.24)}$$

The $n$-th level modified functional integral is defined to be:

$$Z^{(n)}=\int\!\!\exp\{{\imath\over\hbar}[W^{(n)}(\Gamma^{(n)};\hbar ;\hbar)
+G^{(n)}_a(\Gamma^{(n)};\hbar)\pi^{(n)a}]-H^{(n)}(\Gamma^{(n)};\hbar)\}
d\mu^{(n)}, \eqno{(3.25)}$$
where the action $W^{(n)}(\Gamma^{(n)};\hbar;\hbar^{(n)})$ was defined  above,
the final functions $G^{(n)}_a(\Gamma^{(n)};\hbar)$ and
$H^{(n)}(\Gamma^{(n)};\hbar)$ are subordinated by hand to satisfy the
equations obtained from the ones (3.18) -- (3.20), (3.23) by making formal
replacement $n-1\rightarrow n$, the measure $d\mu^{(n)}$ is defined
recursively as follows :

$$d\mu^{(n)}=d\mu^{(n-1)}d\pi^{*(n-1)}d\pi^{(n)},\quad n\ge2, \eqno{(3.26)}$$

$$ d\mu^{(1)}\equiv d\mu. \eqno{(3.27)}$$

By making use of the transformations

$$\delta\Gamma^{(n)A_{(n)}}=(\Gamma^{(n)A_{(n)}},-W^{(n)}+
G^{(n)}_a\pi^{(n)a}+\imath\hbar H^{(n)}){\imath\over2\hbar}\delta X^{(n)}
+{1\over2}(\Gamma^{(n)A_{(n)}},\delta X^{(n)}), \eqno{(3.28)}$$

$$\delta\pi^{(n)a}=[U^{(n)a}_{bc}\pi^{(n)c}
\pi^{(n)b}(-1)^{(\varepsilon_b+n)}
+2\imath\hbar V^{(n)a}_b\pi^{(n)b}+2(\imath\hbar)^2
\tilde{G}{}^{(n)a}]{\imath\over2\hbar}\delta X^{(n)}+\delta\Lambda^{(n)a}_b
\pi^{(n)b}, \eqno{(3.29)}$$
where $\delta X^{(n)}$ is an arbitrary Fermion function, we induce the
following
variations in the integrand of (3.25) :

$$\delta G^{(n)}_a=(G^{(n)}_a,\delta X^{(n)})+
G^{(n)}_b\delta\Lambda^{(n)b}_a, \eqno{(3.30)}$$

$$\delta H^{(n)}=-\Delta^{(n)}\delta X^{(n)}+(H^{(n)},\delta X^{(n)})+
(-1)^{(\varepsilon_a+n)}\delta\Lambda^{(n)a}_a \eqno{(3.31)}$$

The equations for $H^{(n)}$ retain their form under the variations (3.30),
(3.31) by inducing the following transformations for structure coefficients :

$$\begin{array}{c}
\delta U^{(n)a}_{bc}=(U^{(n)a}_{bc},\delta X^{(n)})+
[(\delta\Lambda^{(n)a}_b,G^{(n)}_c)+U^{(n)a}_{bd}
\delta\Lambda^{(n)d}_c]-[(\delta\Lambda^{(n)a}_c,G^{(n)}_b)+\\[9pt]
+U^{(n)a}_{cd}\delta\Lambda^{(n)d}_b](-1)^{(\varepsilon_b+n)(\varepsilon_c+n)}-
\delta\Lambda^{(n)a}_dU^{(n)d}_{bc},
\end{array} \eqno{(3.32)}$$

$$
\delta V^{(n)a}_b=(V^{(n)a}_b,\delta X^{(n)})+V^{(n)a}_d\delta\Lambda^{(n)d}_b-
\delta\Lambda^{(n)a}_dV^{(n)d}_b+[\Delta\delta\Lambda^{(n)a}_b+
(H^{(n)},\delta\Lambda^{(n)a}_b)](-1)^{(\varepsilon_a+n)}
\eqno{(3.33)}$$

$$\delta\tilde{G}{}^{(n)a}=(\tilde{G}{}^{(n)a},\delta X^{(n)})-
\delta\Lambda^{(n)a}_b\tilde{G}{}^{(n)a}. \eqno{(3.34)}$$
By the same reasoning as the one applied to (2.5), we conclude that the
functional integral (3.25) does not depend on the final hypergauge functions
$G^{(n)}_a$. Choosing then $G^{(n)}_a$ to take the simplest form :

$$G^{(n)}_a=\pi^{*(n-1)}_a, \eqno{(3.35)}$$
we obtain

$$Z^{(n)}=Z^{(n-1)},\quad n\ge2, \eqno{(3.36)}$$
and thus arrive at the final reduction formula

$$Z^{(n)}=Z^{(1)}\equiv Z. \eqno{(3.37)}$$

\section{Conclusion}

So, we have constructed the multilevel field-antifield formalism
with the most general Lagrangian hypergauges.

The characteristic feature of the new  generalization  suggested is the
appearance of extra measure factors in the corresponding functional integrals.
It is established that these extra factors can be included naturally into the
multilevel scheme by modifying the boundary conditions to the quantum master
equation.

At the first level it appears to be just possible to construct the extra
measure factor $\exp(-H)$ in an explicitly covariant form [6], that is

$$\exp(-H)=[JM^{-1}\hbox{Ber}(F,G)]^{1/2}, \eqno{(4.1)}$$
where $G_a$ satisfy (2.8), $F^a$, $a=1,\ldots,N$, $\varepsilon(F^a)=
\varepsilon_a+1$, are some functions such that the replacement $\Gamma^A
\rightarrow\bar{\Gamma}{}^A\equiv\{F^a;G_a\}$ is an invertible
reparametrization whose Jacobian is denoted by $J$,

$$J\equiv\hbox{Ber}\left({\partial\bar{\Gamma}\over\partial\Gamma}\right).
\eqno{(4.2)}$$

It is an important property that the expression (4.1), being taken on the
hypergauge surface, does not depend on $F^a$. One can also show that this
expression represents the general solution to the equation (2.9).

Substituting (4.1) into (2.5) we obtain the following representation for the
first level functional integral :

$$Z=\int\!\!\exp({i\over\hbar}W)\delta(G)[JM\hbox{Ber}(F,G)]^{1/2}d\Gamma.
\eqno{(4.3)}$$

Due to the presence of the square root, $[\quad]^{1/2}$, complete measure
factor, entering the integrand of (4.3), cannot be, in general, parametrized
by means of the integral over new fields with a local action.

As for the higher level case, the situation seems to be more complicated.
Regrettably, we are unable, for the present, to construct the corresponding
extra measure factors $\exp(-H^{(n)})$, $n\ge2$, in an explicity covariant
closed form.

{\bf Acknowledgement.} We are thankful to Dr. O.M.Khudaverdian for informing
us on the representation (4.1) prior to its publication.

\end{document}